\documentclass{article}
\usepackage{hq2kpro}
\usepackage{color}

\def \eg      {{\it{e.g.}}}

\def \Dstarpl {D^{\star +}}
\def \Dstar   {D^{\star}}
\def \Dzero   {D^0}

\def \Ds      {D_s}
\def \D       {D}

\def \B       {B}
\def \Bplus   {B^+}
\def \Bs      {B_s}
\def \Bd      {B_d}
\def \Bzs     {B^0_s}
\def \Bzd     {B^0_d}
\def \Bzq     {B^0_q}

\def \Bzdbar  {\overline{B^0_d}}
\def \Bzqbar  {\overline{B^0_q}}

\def \Dmd     {\Delta m_d}
\def \Dms     {\Delta m_s}
\def \Dmq     {\Delta m_q}

\def \micr    {\mu{\mathrm{m}}}
\def \gevc    {{\mathrm{GeV}}/c}
\def \ips     {{\mathrm{ps}}^{-1}}
\def \ps      {{\mathrm{ps}}}
\def \ord     {{\cal{O}}}
\def \amp     {{\cal{A}}}

\def \stat    {_{\mathrm{stat}}}
\def \syst    {_{\mathrm{syst}}}

\begin{document}
\title{REVIEW OF EXPERIMENTAL RESULTS \\ ON
NEUTRAL B MESON OSCILLATIONS} 
\author{Duccio ABBANEO \\
{\em CERN, CH-1211, Geneva 23, Switzerland}
}
\maketitle
\baselineskip=11.6pt
\begin{abstract}
The current status of the experimental knowledge of
neutral B meson oscillations is reviewed. 

The $\Bzd$ oscillation frequency is precisely measured 
by SLD, CDF, and the LEP experiments. 
An overview of the analyses and their combination 
is presented. Preliminary measurements and perspectives at the running 
B factories are also briefly discussed.

The much faster $\Bzs$ oscillations have not yet been resolved,
despite the progress recently achieved
by SLD and ALEPH.
The world combination is presented, together with the expected and observed
lower limit on the $\Bzs$ oscillation frequency. 
The ``amplitude method'', used to combine the analyses in order 
to set the limit, is discussed also as a tool to establish 
the significance of a possible signal.

\end{abstract}
\baselineskip=14pt
\section{Introduction}
\begin{figure}[htb!]
 \vspace{1.6cm}
  \includegraphics{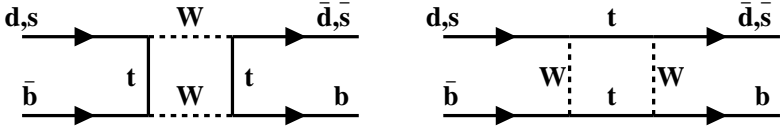}
 \caption{\it
      Box diagrams giving rise to neutral $\B$ meson oscillations.
    \label{fig:fey} }
\end{figure}

One of the main goals of heavy flavour physics is to 
improve our knowledge of quark mixing and establish
whether the Standard Model can describe precisely CP
violating phenomena. Quark mixing is
given by the CKM matrix, which can be written using the
Wolfensteing parameterisation in terms of the three parameters
$\lambda, A, \rho$ and the CP violating phase $\eta$:
\begin{equation}
\left(
\begin{array}{ccc}
V_{ud} & V_{us} & V_{ub} \\
V_{cd} & V_{cs} & V_{cb} \\
V_{td} & V_{ts} & V_{tb} 
\end{array}
\right) \approx 
\left(
\begin{array}{ccc}
1-\frac{1}{2}\lambda^2 & \lambda               &A\lambda^3(\rho- i \eta) \\
-\lambda               &1-\frac{1}{2}\lambda^2 & A\lambda^2 \\
A\lambda^3(1-\rho- i \eta) & -A\lambda^2       & 1
\end{array}
\right) \ .
\label{eq:CKM}
\end{equation}
Neutral $\B$ meson oscillations are described
by the ``box diagrams'' (Fig.~\ref{fig:fey}). 
The oscillation frequency in the $\Bzd - \Bzdbar$ system,
which is proportional to the mass difference of the two
eigenstates, can be translated into a measurement of $\mid V_{td} \mid$,
and therefore it yields information on the CP violating phase $\eta$
(see Eq.~\ref{eq:dmd}).
Unfortunately QCD effects are large and the associated 
uncertainty dominates the extraction of  $\mid V_{td} \mid$.
\begin{equation}
  \Dmd \propto {\mid  V_{td} \mid}^2 \cdot {\cal{F}}({\mathrm{QCD}}) \ .
  \label{eq:dmd}
\end{equation}
A better constraint on $\eta$ could be obtained from the ratio of
the oscillation frequencies of $\Bs$ and $\Bd$ mesons, since some
of the QCD uncertainties cancel in the ratio. The factor
$\xi$ in Eq.~\ref{eq:ratio} is estimated to be known at the $5\%$ level.
\begin{equation}
  \frac{\Dms}{\Dmd} = 
  \frac{m_{\Bs}}{m_{\Bd}} \xi^2 {\left| \frac{V_{ts}}{V_{td}} \right|}^2 \ .
  \label{eq:ratio}
\end{equation}
The
proper time distributions of ``mixed'' and ``unmixed'' decays, given in Eq.~\ref{eq:pdf},
are measured experimentally.
The oscillating term introduces a time-dependent difference between the two classes.
\begin{eqnarray}
  {\cal P}(t)_{\Bzq \to \Bzqbar} & = & 
  \frac{\Gamma e^{-\Gamma t}}{2} \, [1 - \cos (\Dmq \, t)] \ ,
  \nonumber \\
  {\cal P}(t)_{\Bzq \to \Bzq} & = & 
  \frac{\Gamma e^{-\Gamma t}}{2} \, [1 + \cos (\Dmq \, t)] \ , 
  \label{eq:pdf}
\end{eqnarray}
The amplitude of such difference is damped not only by the 
natural exponential decay,
but also by the effect of the experimental resolution in the proper time
determination. The proper time is derived from the measured decay length
and the reconstructed momentum of the decaying meson. The resolution on the decay length
$\sigma_L$
is to first order independent of the decay length itself, and is largely determined
by the tracking capabilities of the detector. The momentum resolution $\sigma_p$
depends widely on 
the final state chosen for a given analysis, and is typically proportional to the momentum 
itself. 
\begin{figure}[bt!]
  \vspace{7.2cm}
  \includegraphics{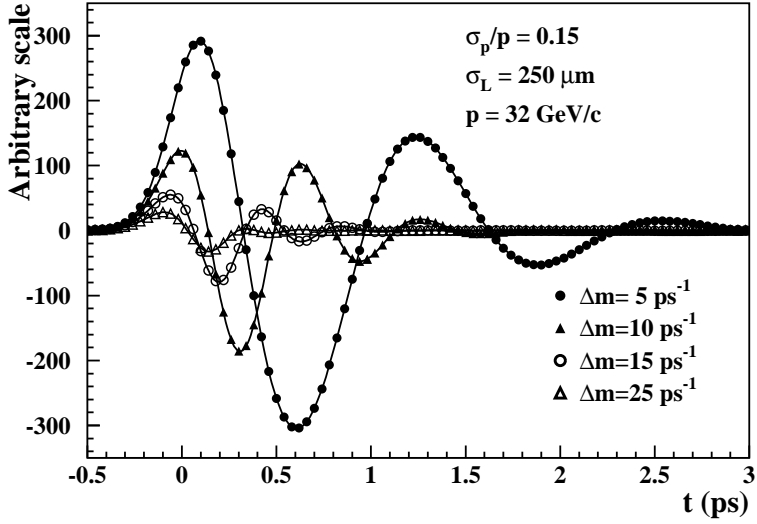}
  \caption{\it
    Difference in the proper time distributions of unmixed and mixed decays
    for monochromatic $\B$ mesons, fixed
    decay length and momentum resolutions, and different values of the
    oscillation frequency.
    \label{fig:time} }
\end{figure}
The proper time resolution can be therefore written as:
\begin{equation}
  \sigma_t = \frac{m}{p} \sigma_L \oplus \frac{\sigma_p}{p} \, t \ ,
  \label{eq:timeres}
\end{equation}
where the decay length resolution contributes a constant term, and the momentum resolution
a term proportional to the proper time. Examples of the resulting difference are 
shown in Fig.~\ref{fig:time}, for the simple case of monochromatic $\B$ mesons of 
momentum $32 \ \gevc$, resolutions of $\sigma_p/p = 0.15$ and $\sigma_L =250\
\micr$ (Gaussian), and for different values of the true oscillation frequency.
For low frequency several periods can be observed. As the frequency 
increases, the
effect of the finite proper time resolution becomes more relevant, 
inducing an overall decrease of observed difference, and a faster damping as a function
of time (due to the momentum resolution component). 
In the example given for a frequency
of $25\ \ips$ only a small effect corresponding to the
first half-period can be seen.

\section{Analysis methods}

The first step for a  $\B$ meson oscillation analysis is the selection of
final states suitable for the study. The choice of the selection criterion
determines also the strategy for the tagging
of the meson flavour at decay time. Then, the flavour at production time
is tagged, to give the global mistag probability

Finally, the proper time is reconstructed for each meson candidate, 
and the oscillation is studied by means of a likelihood fit to the 
distributions of decays tagged as mixed or unmixed.
The different selection methods and the techniques to tag the flavour 
at production time
are briefly discussed below.
Up-to-date references for all the $\Dmd$ and $\Dms$ analyses
can be found at 
\begin{center}
{\tt{http://lepbosc.web.cern.ch/LEPBOSC/}}.
\end{center}

\subsection{Selection methods}
\label{sec:sele}

A variety of selection methods have been developed for $\Bd$ and $\Bs$
oscillation analyses, which offer different advantages in terms
of statistics, signal purity and resolution.

Fully inclusive selections are used by SLD for both $\Dmd$ and $\Dms$ analyses.
At LEP they have been attempted by ALEPH for $\Dmd$ and DELPHI for $\Dms$.
These methods yield very high statistics ($\ord(10^5)$ decays at LEP, $\ord(10^4)$
at SLD). The signal fraction is what is given by nature ($\approx 40\%$ for $\Bd$ and
$\approx 10 \%$ for $\Bs$). There is no
straightforward method for tagging the flavour 
at decay time. At LEP many different variables are combined by means of a neural network,
while at SLD the excellent tracking capabilities allowed the development of the
dipole technique, discussed later in Section~\ref{sec:sldip}.

Semi-inclusive selections rely on the identification of one specific 
$\B$ meson decay product. This is typically a lepton, (electron or muon),
but it can be a fast kaon in the case of  $\Dmd$ analyses (used by SLD).
The charge of this particle gives the flavour at decay time, with small
mistag probability. As for fully inclusive methods, there is no
enhancement in signal fraction from the selection, 
and statistics are typically a factor of ten smaller.
These methods are used by all experiments.
In some cases an attempt is made to select a second $\B$ meson decay product 
in order to enhance the signal fraction. This is typically a ``soft'' pion
from the decay $\Dstarpl \to \Dzero \pi^+$ for $\Dmd$ analyses, a kaon with charge 
opposite to the lepton, or a $\phi$, in the case of $\Dms$ analyses.

In semi-exclusive selections the charmed meson from the $\B$ meson decay
is fully reconstructed: a $\D$ or $\Dstar$ for  $\Dmd$ analyses,
a $\Ds$ for  $\Dms$ analyses. The charmed meson reconstruction 
can be combined with the tagging
of the lepton from the $\B$ meson decay, in which case only the neutrino
is undetected. These methods have substantially lower statistics, but the
signal purity goes up to $60\%$ and the final state mistag probability is very small.

At LEP, ALEPH and DELPHI have also attempted full reconstruction of $\Bs$ decays.
The selections yield only about 50 candidates, with $\approx 50\%$ estimated 
signal purity.
Nevertheless, since all the decay products are identified, these events have excellent
proper time resolution, and therefore give useful information in the high frequency range.

\subsection{Flavour tagging at production time}

All particles of the event except those tagged as the $\B$ meson 
decay products can be used to derive information on the $\B$ meson  
flavour at production time.

In the hemisphere containing the B meson candidate, charged particles 
originating from the primary vertex, produced in the hadronization of the
$b$ quark, may retain some memory of its charge. Either the charges of all 
tracks are combined, weighted according to the track kinematics, or the track
closer in phase space to the  $\B$ meson candidate is selected, requiring that it be
compatible with a kaon for  $\Dms$ analyses, with a pion for $\Dmd$ analyses.

In the opposite hemisphere, the charge of the other $b$ hadron can be tagged, exploiting
the fact that $b$ quarks are produced in pairs. Tracks from the $b$-hadron decay 
can be distinguished from fragmentation tracks on a statistical basis, from their
compatibility with the primary vertex or with an inclusively-reconstructed secondary 
vertex. Inclusive hemisphere-charges can be formed assigning weights to  tracks according to the
probability that they belong to the primary or the secondary vertex, complemented with
more ``traditional'' jet-charges, where track weights are defined on the basis of track kinematics.
Furthermore, specific decay products, such as leptons or kaons, 
can be searched for also in the opposite hemisphere, and their charge used 
as an estimator of the quark charge.

Finally, both at LEP and at SLD, the polar angle of the $\B$ meson candidate 
is also correlated with the charge of the quark, because of the 
forward-backward asymmetry in the Z decays. This correlation is particularly
relevant at SLD, 
due to the polarisation of the electron beam.

\begin{figure}[tb!]
  \vspace{5cm}
  \includegraphics{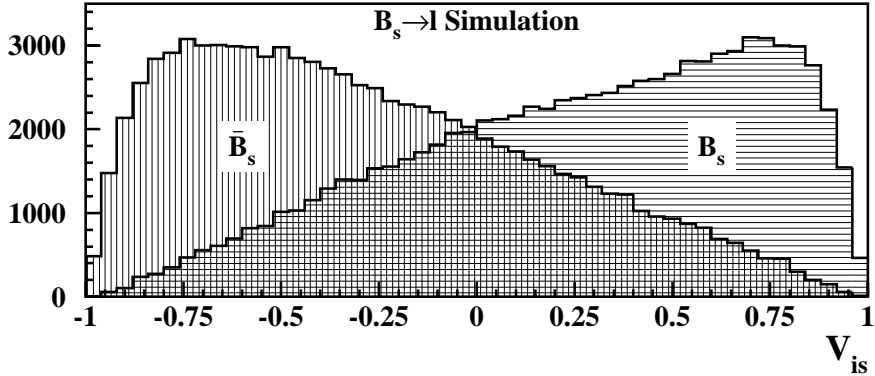}
  \caption{\it
    Variable for flavour tagging at production time, 
for the inclusive lepton $\Dms$ analysis of ALEPH. 
Distribution of simulated signal events.
    \label{fig:is} }
\end{figure}

Many of the production-flavour estimators mentioned above are correlated
among themselves. The most recent and sophisticated analyses have attempted to
use efficiently all the available information by combining the different estimators
using neural network techniques. In Fig.~\ref{fig:is} the combined initial state tag
variable $V_{\mathrm{is}}$ is shown, 
for the case of  the recent $\Dms$ analysis with inclusive lepton 
tag from ALEPH. The value of $V_{\mathrm{is}}$ is translated into
a mistag probability, evaluated event-by-event. 
The lowest values for the production flavour mistag obtained at LEP are just below
$25\%$, while at SLD values smaller than $15 \%$ are achieved. The difference is
mostly due to the beam polarisation.

\boldmath
\section{Measurements of $\Bd$ oscillations}
\unboldmath

\begin{figure}[tb!]
 \vspace{15cm}
  \includegraphics{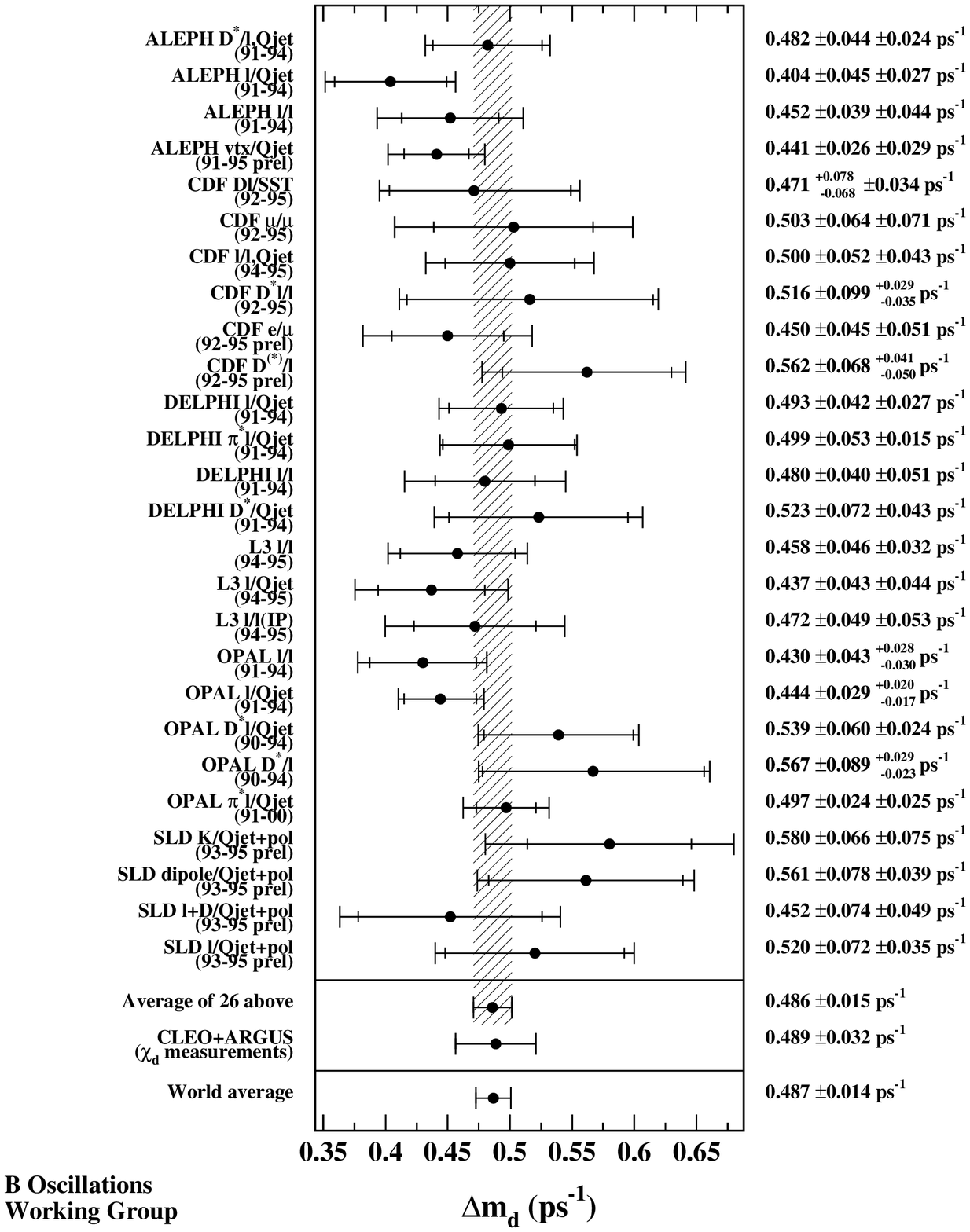}
 \caption{\it
      Summary of $\Dmd$ analyses from  SLD, CDF, and the LEP experiments,
together with the current world average (excluding asymmetric $\B$ factories).
    \label{fig:bd_world} }
\end{figure}
A large number of analyses have been developed by  SLD, CDF, and the LEP experiments,
using the different selection techniques outlined in Section~\ref{sec:sele}.
They are summarised in Fig.~\ref{fig:bd_world}.
One of the most recent and precise is the OPAL preliminary analysis based
on inclusive reconstruction of $\Bd$ semileptonic decays.
\begin{figure}[b!]
 \vspace{10cm}
  \includegraphics{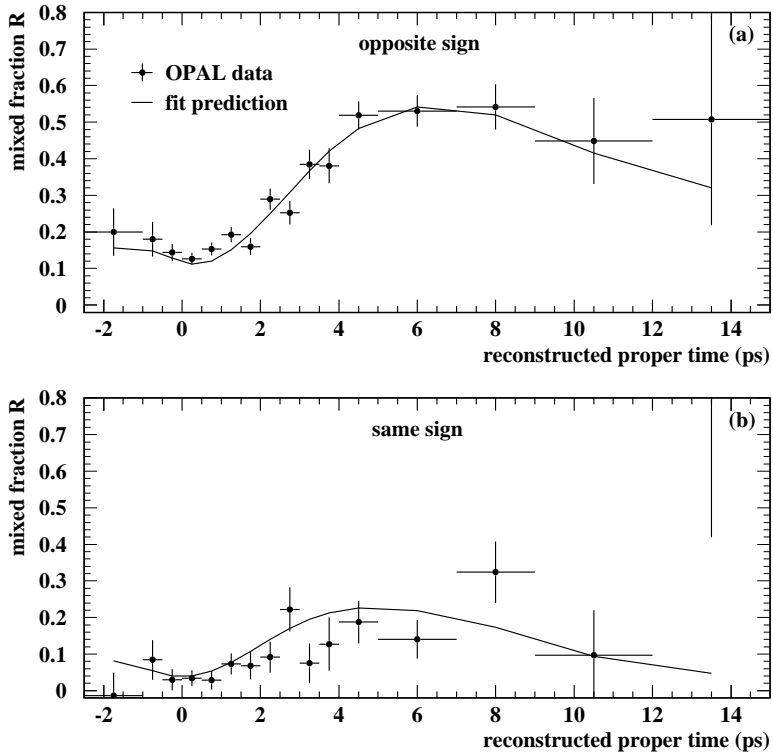}
 \caption{\it
   Fraction of events tagged as mixed as a function of the reconstructed
   proper time, for the correct (a) and wrong (b) charge combinations. 
   Superimposed is the curve predicted for the fitted value of $\Dmd$.
    \label{fig:opal} }
\end{figure}
The selection aims at tagging $\Bd \to \Dstarpl \ell^- \bar \nu$ decays, with
$\Dstarpl \to \Dzero \pi_{\mathrm{s}}^+$. The $\Dzero$ is reconstructed inclusively
by selecting tracks and neutral objects compatible with the $\Dzero$ kinematics.
The $\Bd$ decay vertex is reconstructed by intersecting the lepton with the soft pion.
The neutrino energy is derived from the measured missing energy.
The momentum resolution achieved is about $10\%$, 
with a mistag probability of about $28\%$. Combinations where the lepton and the soft pion
have the same charge provide a useful control sample enriched in combinatorial background.
The fraction of events tagged as mixed is displayed as a function of the reconstructed
proper time in Fig.~\ref{fig:opal}, for both charge combinations, 
together with the curve predicted for the fitted value of $\Dmd$.
The analysis of all the available $Z$ peak statistics, including those 
collected in the years 1996-2000 for the detector calibration, yields:
\begin{equation}
\Dmd = 0.497 \pm 0.024 \stat \: \pm 0.025 \syst  \: \ \ips \ .
\label{eq:opaldmd}
\end{equation}

\subsection{The averaging method}

All the $\Dmd$ measurements reported in Fig.~\ref{fig:bd_world} require
to some extent input from simulated events.
Quantities derived from the
simulation are affected by uncertainties on the physics processes that are 
simulated. These sources are
common to all measurements, and they have to be treated as correlated
when averaging individual results. Furthermore, in order to produce consistent
measurements, and hence meaningful averages, the input physics parameters
used in the simulation must be the same for all analyses in all
experiments. For these reasons, all results are first adjusted 
to a common set of relevant input parameters, (\eg\ $b$-hadron lifetimes and production
fractions). Then results from  SLD, CDF, and the LEP experiments are averaged, obtaining
a first value for $\Dmd$. Following that, time-integrated measurements of 
$\chi_d$ from experiments at the $\Upsilon {\mathrm{(4s)}}$ resonance are considered, 
and translated to $\Dmd$ measurements. Measurements of the average time-integrated mixing parameter at the $Z$ energy ($\bar \chi = f_{\Bs} \chi_s + f_{\Bd} \chi_d$) are also considered 
and translated to a constraint for the $\Bs$ production rate at the $Z$. This yields
a new value of $\Dmd$ and new $b$-hadron production fractions, which are used 
for a further iteration. The procedure is repeated until convergence.

The method yields a world average of $\Dmd$ and consistent world average values
for the $b$-hadron production fractions:
\begin{equation}
f_{\Bd , \Bplus} = (40.0 \pm 1.0) \% \ \ \  \ \ 
f_{\Bs} = (9.7 \pm 1.2) \%  \ \ \  \ \ 
f_{\mathrm{baryon}} = (10.3 \pm 1.7) \%  \ .
\end{equation}
\begin{equation}
\Dmd  =  0.487 \pm 0.014 \ \ips \ . 
\end{equation}

\subsection{The new results from the asymmetric $\B$ factories}

\begin{figure}[tb!]
 \vspace{8cm}
  \includegraphics{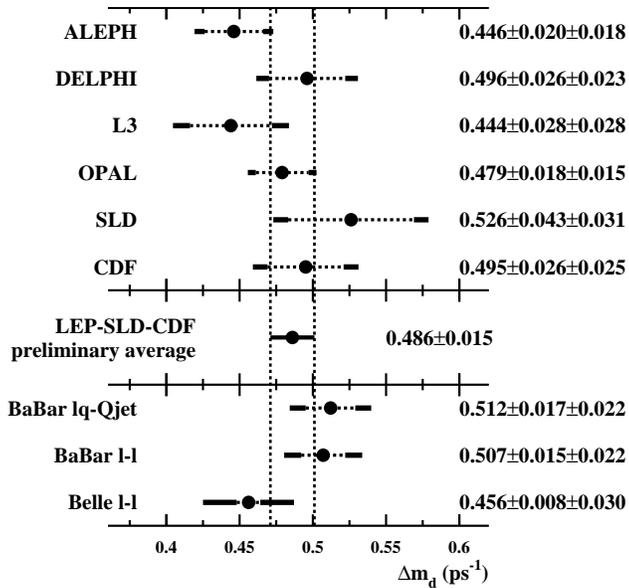}
 \caption{\it
       $\Dmd$ averages from  SLD, CDF, and the LEP experiments, compared 
to the new results from the asymmetric $\B$ factories.
    \label{fig:mybd} }
\end{figure}

New preliminary results from the asymmetric $\B$ factories have been made
public this year. They are compared in Fig.~\ref{fig:mybd}
to the averages of the other experiments. The two
BaBar results should not be averaged since they have some statistical and
systematic correlation that has not been estimated.
The asymmetric $\B$ factories have already exceeded the statistical 
precision of the world average. At the moment they are quoting rather large
systematic errors, which are however dominated by uncertainties on 
parameters that are also measured in their data. Therefore the large statistics
expected in the near future will allow very precise measurements to be made.

\boldmath
\section{Studies of $\Bs$ oscillations}
\unboldmath

The analyses completed so far are not able to resolve 
the fast $\Bs$ oscillations: they can only exclude a 
certain range of frequencies. Combining such excluded ranges
is not straightforward, and a specific method, called ``amplitude method''
has been introduced for this purpose\cite{hgm}. In the likelihood fit to 
the proper time distribution of decays tagged as mixed or unmixed, the 
frequency of the oscillation is not taken to be the free parameter,
but it is instead fixed to a ``test'' value $\omega$. An auxiliary 
parameter, the  amplitude $\amp$ 
of the oscillating term is introduced, and left free in the fit.
The proper time distributions for unmixed and unmixed decay, prior to 
convolution with the experimental resolution, are therefore written as
\begin{equation}
  {\cal P}(t) = 
  \frac{\Gamma e^{-\Gamma t}}{2} \, [1 \pm \amp \cos (\omega\, t)] \ ,
\end{equation}
with $\omega$ the test frequency and $\amp$ the only free parameter. 
When the test frequency is much smaller than the true frequency 
(\mbox{$\omega \ll \Dms$}) the expected value for the amplitude is 
\mbox{$\amp = 0$}.
At the true frequency (\mbox{$\omega = \Dms$}) the expectation is 
\mbox{$\amp = 1$}.
All the values of the test frequency $\omega$ for which 
\mbox{$\amp + 1.645 \sigma_\amp < 1$} are excluded at $95\%$~C.L.

The amplitude has well-behaved errors, and 
different measurements can be combined in a straightforward way, 
by averaging the amplitude measured at different test frequencies.
The excluded range is derived from the combined amplitude scan.

\subsection{Examples of $\Dms$ analyses}

Some examples of recent and relevant analyses are presented 
in the following, and their statistical power discussed and compared.

\subsubsection{\it Exclusive $\Bs$ reconstruction at LEP}

\begin{figure}[bt!]
 \vspace{10.7cm}
  \includegraphics{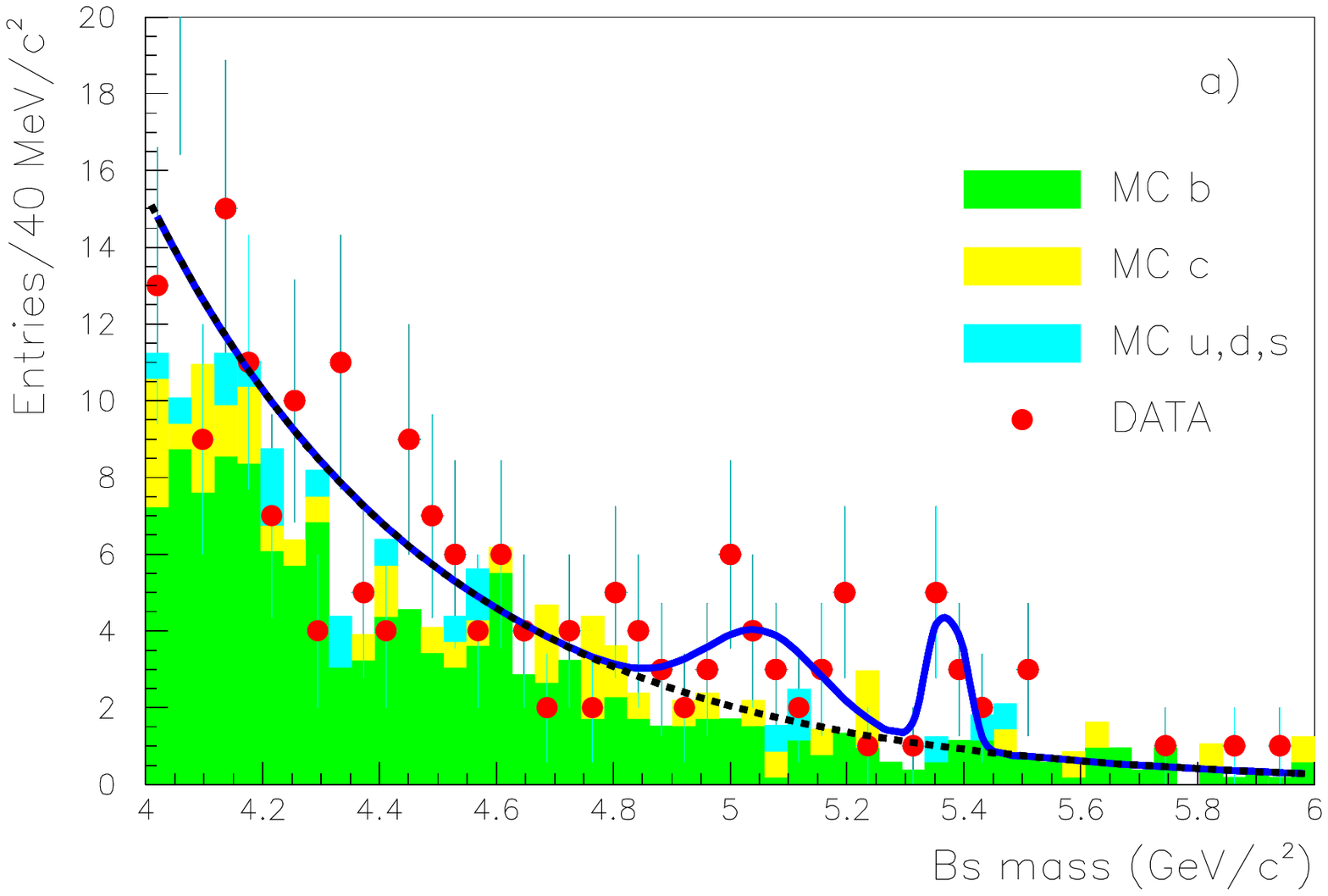} 

  \includegraphics{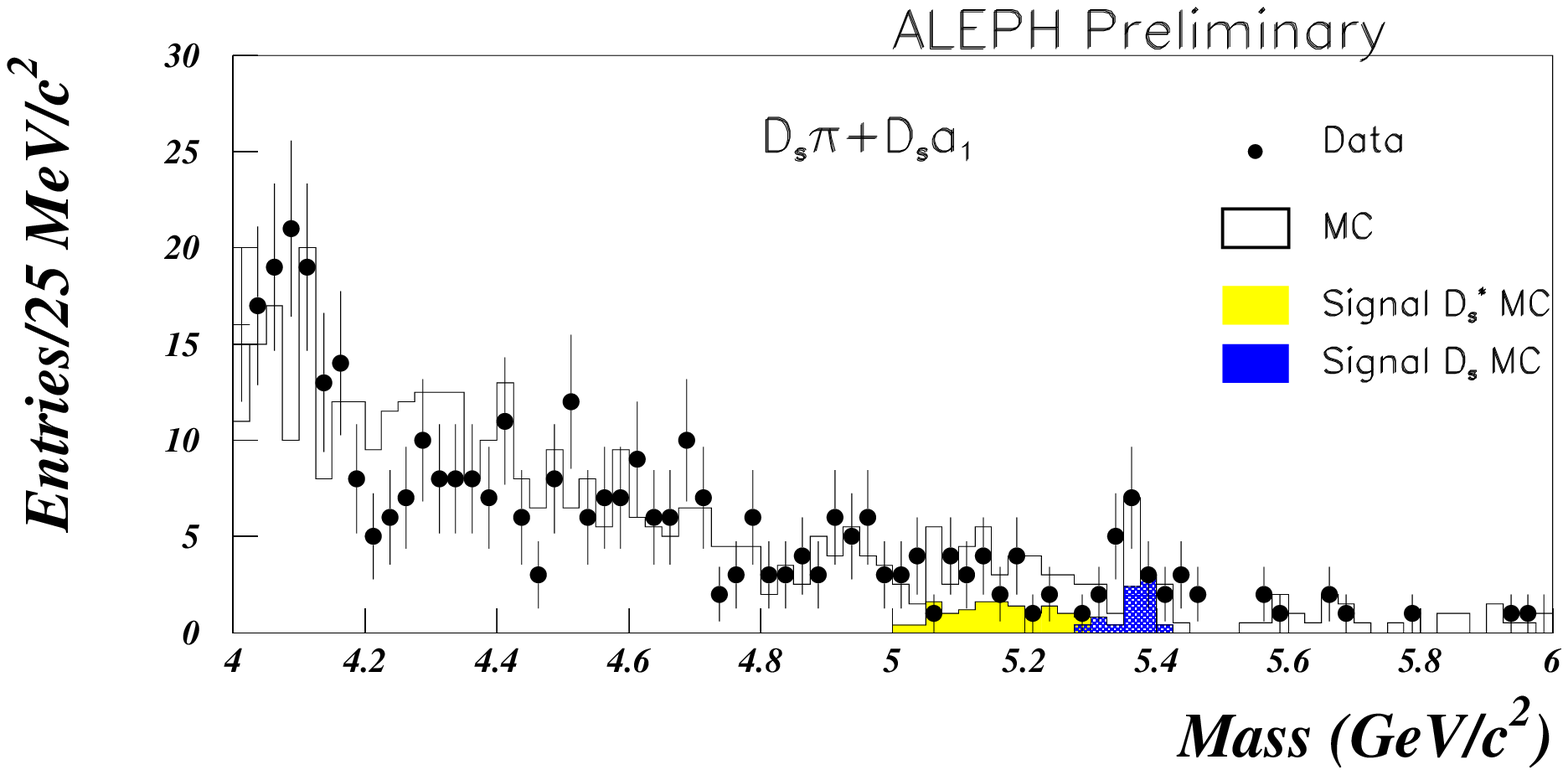}
  \color{white}

  \vspace{-10.3cm}
  \hspace{8.5cm} \rule{1cm}{0.5cm} 

\color{black}
 \hspace{4cm} DELPHI

\vspace{-1.1cm} 

 \hspace{9.1 cm} (a)

\vspace{6.cm}

 \hspace{9.1 cm} (b)

  \vspace{3.5cm}
 \caption{\it
      Reconstructed $\Bs$ mass in DELPHI (a) and ALEPH (b).
    \label{fig:bsexcl} }
\end{figure}
Analyses of fully reconstructed $\Bs$ candidates have been 
performed by ALEPH and DELPHI,
exploiting the decay channels \mbox{$\Bzs \to D_s^{(\star)-} \pi^+$} and 
\mbox{$\Bzs \to D_s^{(\star)-} a_1^+$}. DELPHI also uses 
\mbox{$\Bzs \to \overline{\Dzero} K^- \pi^+$} and
\mbox{$\Bzs \to \overline{\Dzero} K^- a_1^+$}. 
The reconstructed mass structure consists of a 
narrow peak at the nominal $\Bs$ mass 
plus a broad shoulder at lower masses due to the decays
involving a \mbox{$D_s^\star \to D_s \gamma$} with the photon undetected (see Fig.~\ref{fig:bsexcl}).

The signal purity is \mbox{$\approx 50\%$}, with a proper time resolution of 
\mbox{$\sigma_t \approx 0.08 \ \ps$}, to be compared with 
\mbox{$\sigma_t \approx 0.2\div 0.3\ \ps$}
for the other analyses at LEP.

\subsubsection{\it Analyses of $\Ds^- \ell^+$ final states at LEP}

At LEP, these methods represent a good compromise between statistics, 
signal purity and resolution.

\begin{figure}[t!]
 \vspace{4.9cm}
  \includegraphics{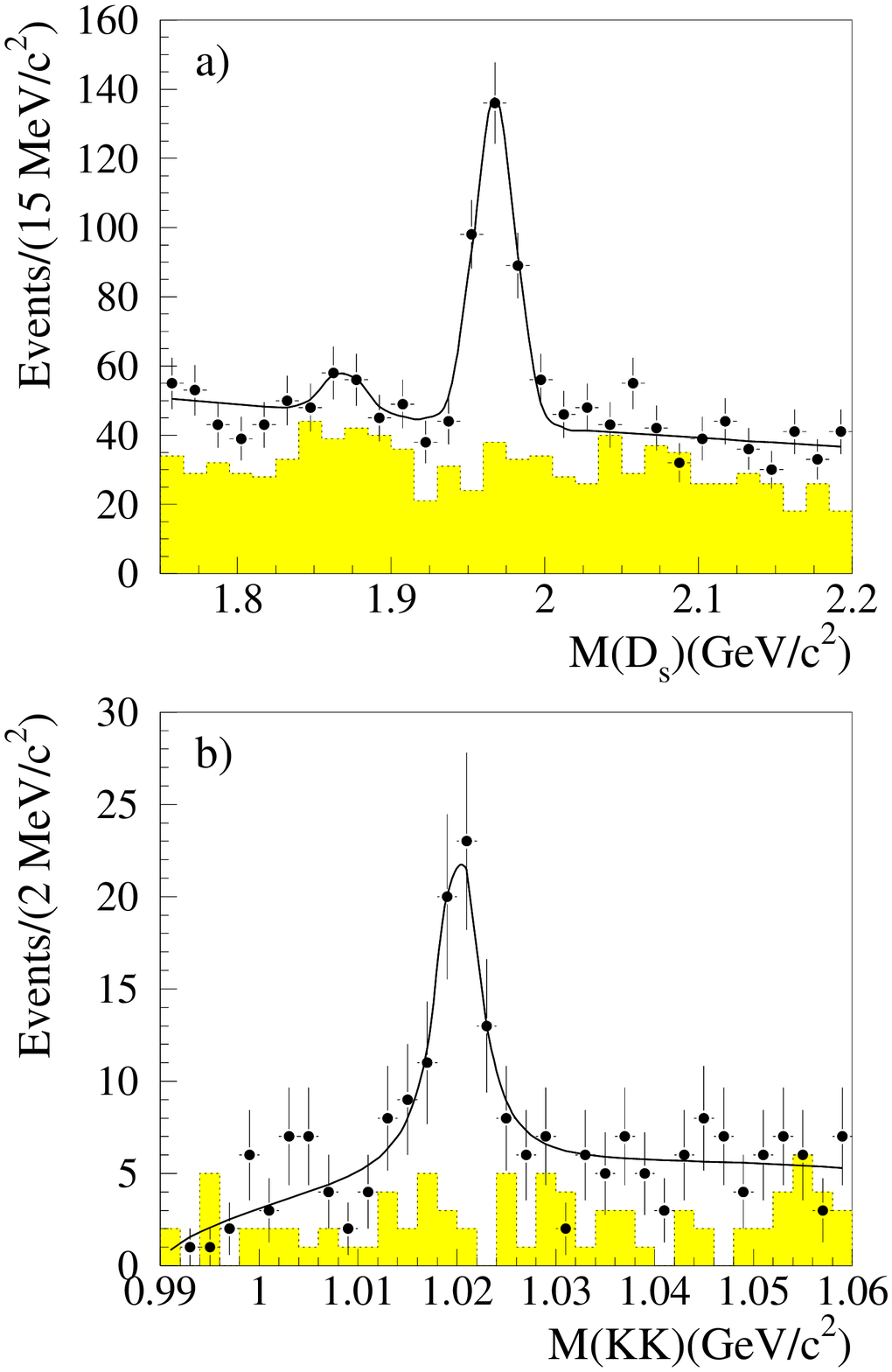}
  \includegraphics{dsl_d.eps}
\color{white}

  \rule[-5.1cm]{3cm}{5.7cm} \hspace{-.5cm} \rule[-5.1cm]{4cm}{5.4cm}

  \vspace{-15.4cm}
  \hspace{6cm} \rule{6.5cm}{5cm}

  \vspace{4.5cm}
\color{black}
 \caption{\it
      Reconstructed $\Ds$ mass peak for the DELPHI $\Ds^- \ell^+$ analysis, for the 
fully hadronic $\Ds$ decay channels (a). For the same analysis, $\phi$ mass peak for
the $\Ds^- \to \phi \ell^-$ channel (b). 
    \label{fig:dsl} }
\end{figure}

The signal purity achieved is typically around or above $50\%$, with 
about $40\%$ combinatorial background and  $10\%$ physical background
from $\B \to \Ds \bar \D$ decays, with $\bar \D \to \ell^- \bar \nu X$.
The combinatorial background is controlled from the data using the $\Ds$
mass peak sidebands (see Fig.~\ref{fig:dsl}).
The decay length resolution is $\approx 200\ \micr$ and the momentum 
resolution about $10\%$ due to the undetected neutrino, whose momentum 
is derived from the measured missing energy. Statistics are typically
a few hundred candidates.

\subsubsection{\it Analyses of inclusive lepton samples at LEP and SLD}
\label{sec:incll}

These analyses give the highest sensitivity both at LEP and at SLD.
Lepton candidates from direct $b$-hadron decays are selected on the basis 
of the lepton kinematics; non-$b$ background is suppressed by applying 
lifetime-based $b$ tagging algorithms.
The charmed particle is reconstructed inclusively 
and intersected with the lepton to find the $b$ decay vertex. 
At LEP, the resolution has typically a core of $250\ \micr$
with tails up to 1~mm, worse than for $\Ds \ell$
final states, due to the contribution of missing or misassigned tracks.
At SLD, the small and precise CCD vertex detector allows a core resolution
of about $60\ \micr$ to be achieved.
The neutrino momentum is estimated from the missing energy, with resolution
similar to the $\Ds \ell$ case.

\begin{figure}[bt!]
 \vspace{7.0cm}
  \includegraphics{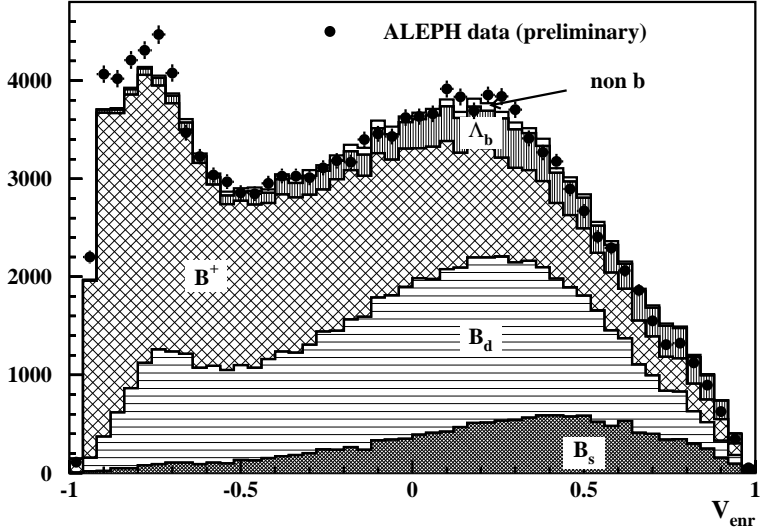}
 \caption{\it
      Variable discriminating $\Bs$ decays from other $b$ hadrons, 
for the inclusive lepton $\Dms$ analysis of ALEPH. 
The signal fraction is evaluated event-by-event 
as a function to such a variable. 
    \label{fig:enr} }
\end{figure}

The sensitivity of the analysis can be further improved by estimating 
the probability that each event be signal, instead of using the 
average $\Bs$ fraction
(about $10\%$). The total charge flowing out of the $b$ vertex 
is a good discriminant between $\Bplus$ and neutral $b$ hadrons. 
The presence of kaons from the primary vertex, or among the tracks 
assigned to the $c$ meson, can discriminate between $\Bs$ mesons 
and other $b$ hadrons. Such information can be combined into a single 
discriminating variable (see Fig.~\ref{fig:enr}) 
and the signal fraction evaluated event by event, 
as a function of such variable. 
This procedure enhances the statistical power 
of the analyses by up to $25\%$.

\subsubsection{\it The fully inclusive analysis at SLD}
\label{sec:sldip}

In a fully inclusive analysis, both the $c$ and the $b$ decay vertices
are reconstructed topologically. This requires excellent tracking
capabilities, to keep the contribution of missing or misassigned
tracks at an acceptable level. The method is particularly suitable for SLD,
where the available statistics are ten times smaller than in a LEP 
experiment, but with much more precise tracking.

\begin{figure}[tb!]
 \vspace{8.1cm}
  \includegraphics{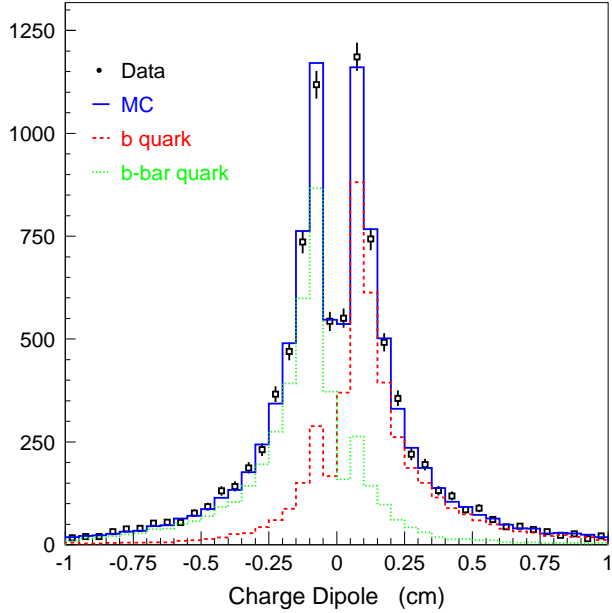}
 \caption{\it
      Flavour tag at decay time for the fully inclusive analysis
of SLD.
    \label{fig:dipole} }
\end{figure}

Since no specific $b$ decay product is tagged, the flavour at decay time
has to be identified with a dedicated method. SLD takes advantage of the 
good separation between the $\B$ and the $\D$ meson vertices 
to build a variable which 
measures the ``charge flow'' between the two, (``charge dipole'') 
which exploits the fact that a $\Bs$ meson 
decays to a $\Ds^-$ meson, while a $\overline{\Bs}$ decays to a $\Ds^+$. 
The kind of separation achieved can be seen in Fig.~\ref{fig:dipole}.

The statistics collected are four times larger than for the inclusive lepton 
analysis, which compensates for the slightly worse decay length resolution
(core of $\approx 70\ \micr$) and the larger dilution in the flavour tag at decay 
time, yielding a very similar statistical power.

\subsubsection{\it Comparison of analyses}

\begin{figure}[htb!]
 \vspace{7.5cm}
  \includegraphics{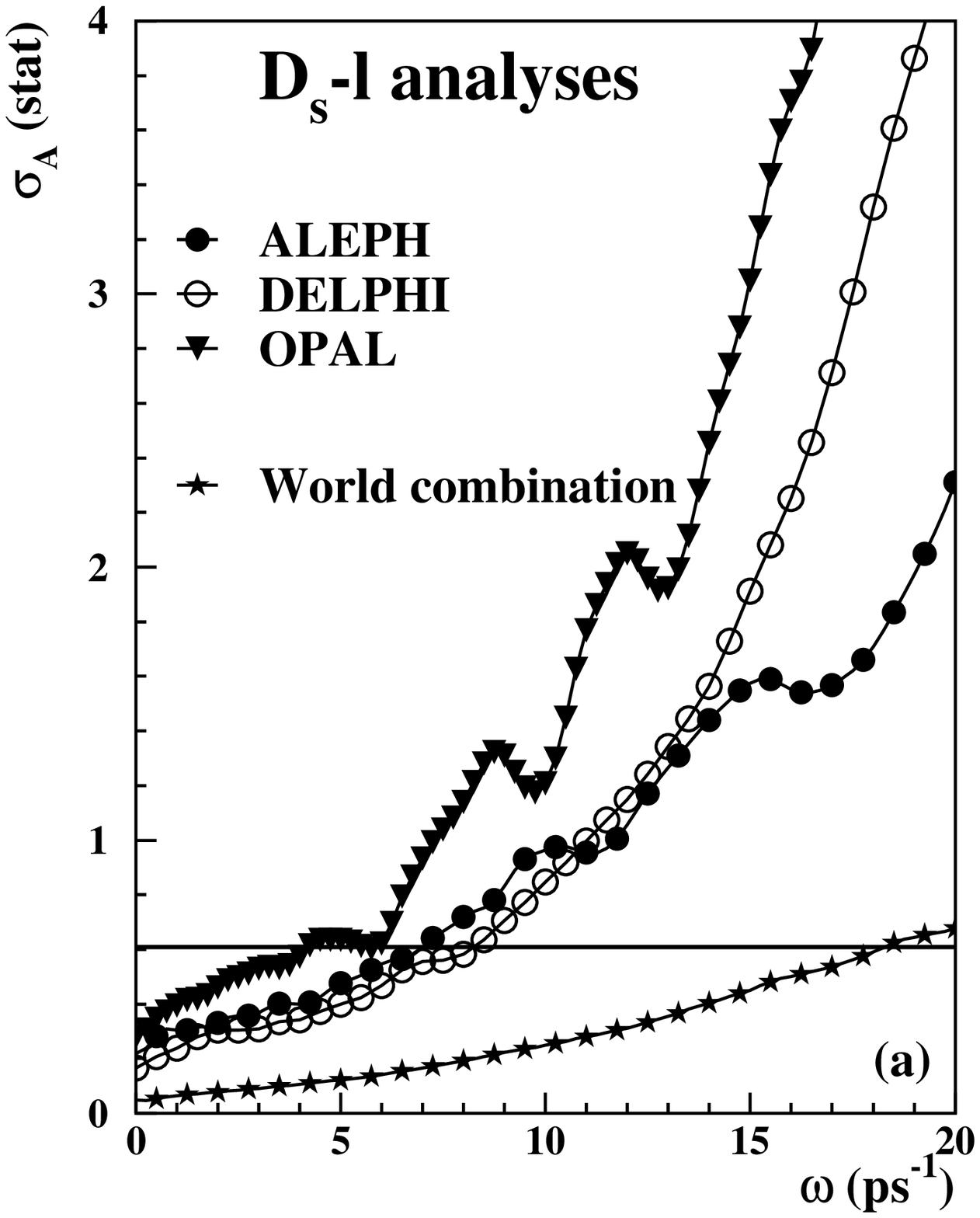}
  \includegraphics{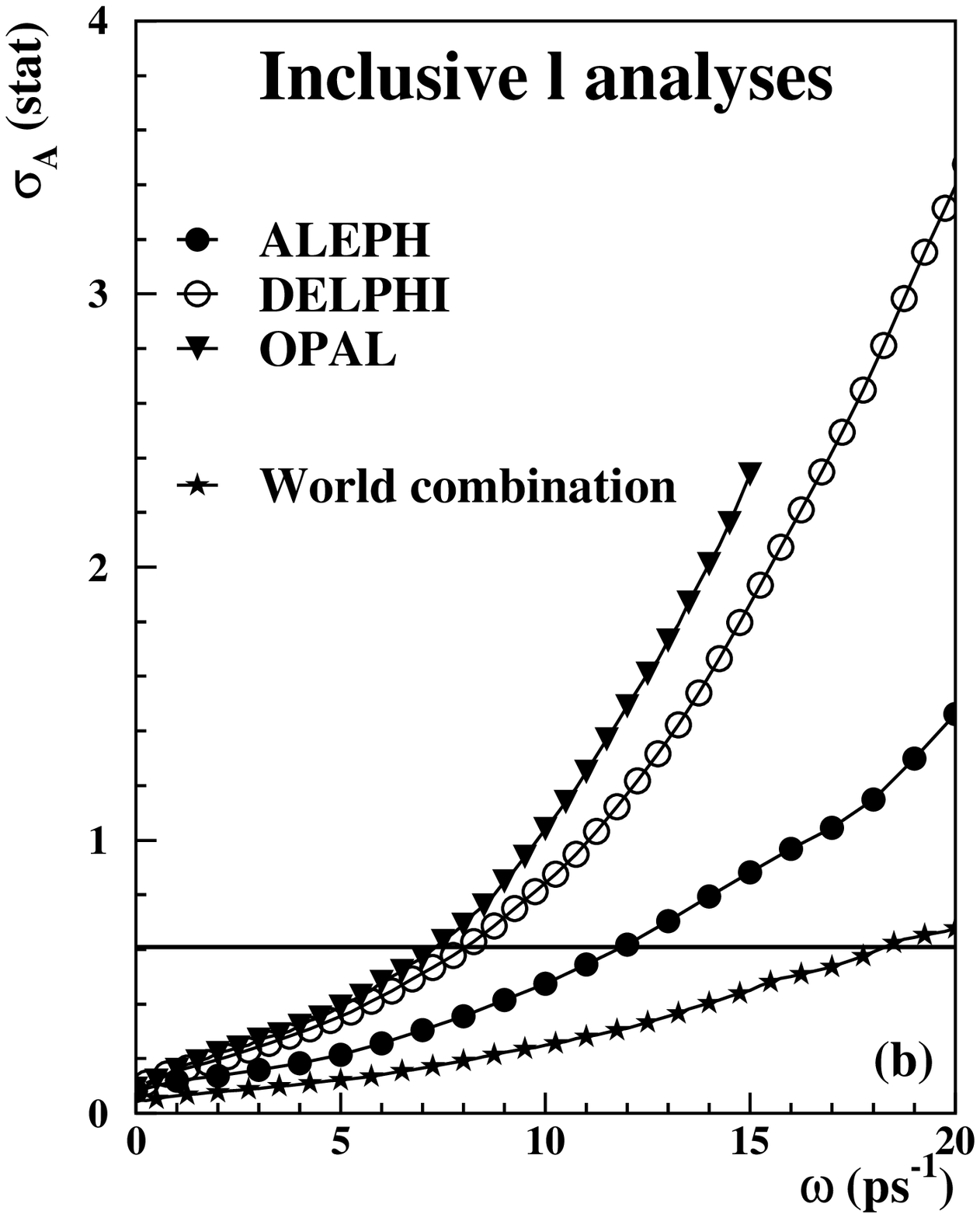}
 \caption{\it
   Statistical error on the measured amplitude as a function of test frequency
for the analyses of $\Ds \ell$ final states (a) and inclusive lepton samples 
(b) at LEP. The world combination of all analyses is shown for comparison.
    \label{fig:complep} }
\end{figure}

The available $\Ds \ell$ and inclusive lepton analyses at LEP are compared
in Fig.~\ref{fig:complep}, for what concerns their statistical power.
\begin{figure}[htb!]
 \vspace{7.5cm}
  \includegraphics{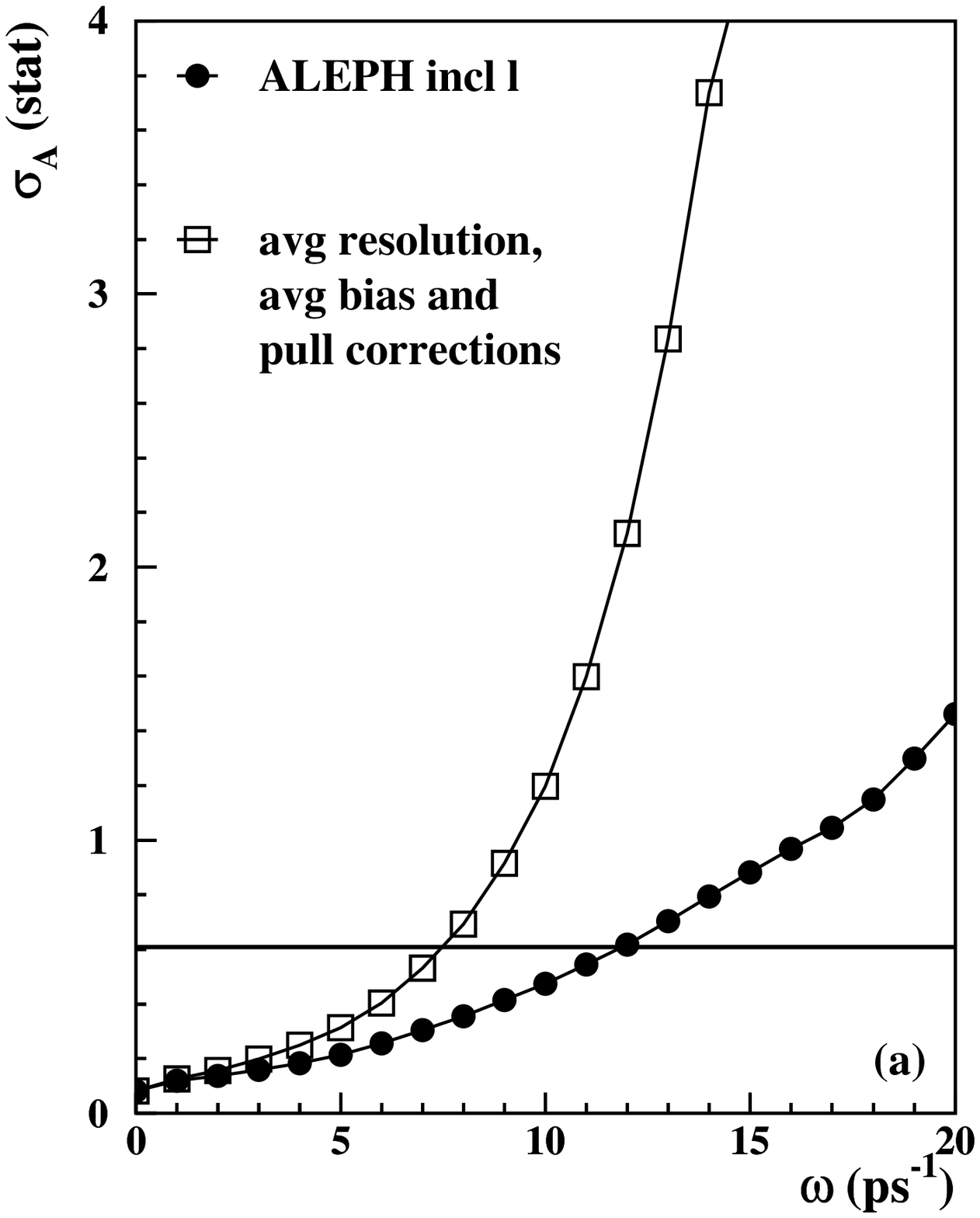}
  \includegraphics{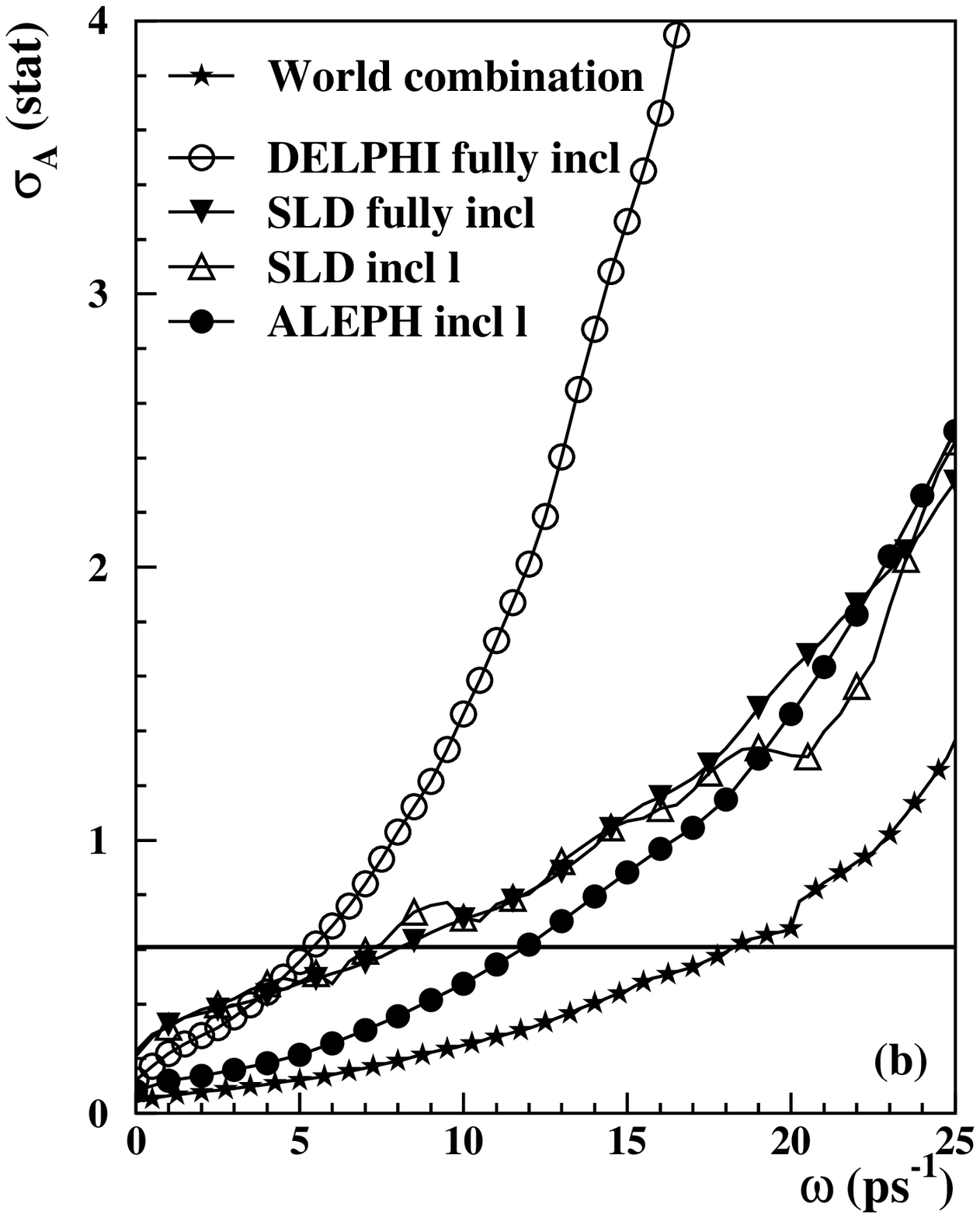}
 \caption{\it
  (a) Comparison of the statistical errors on the amplitude 
for the ALEPH inclusive lepton analysis, with the same analysis if the 
event-by-event treatment of bias and pull corrections and decay length uncertainty is dropped. \newline
(b) Comparison of the SLD inclusive lepton and fully inclusive analyses, with the corresponding best available at LEP.
    \label{fig:comp2} }
\end{figure}

A striking feature is that the  $\Ds \ell$ analysis from DELPHI 
has a slightly smaller statistical error than the ALEPH one at low frequency, 
but it becomes much worse at high frequency (the statistical error is larger
by more than a factor of two).
Similarly the slope of the error curve for the ALEPH inclusive lepton analysis 
is substantially more flat than for DELPHI.

A plausible reason for that is demonstrated in Fig.~\ref{fig:comp2}a, 
where the ALEPH
inclusive lepton analysis is compared with the same analysis if the 
event-by-event treatment of the decay length bias correction
and error assignment is dropped. In the analysis the decay length 
is reconstructed by intersecting the charm meson candidate direction with
the lepton track, and its uncertainty is estimated from the fit error. The 
fitted decay length needs to be corrected for a small bias caused by 
the selection (bias correction), and the fit error must be enlarged to account
for the contribution of missing or misassigned tracks (pull correction).
These corrections, taken from the simulation, depend strongly on the event 
topology: they are negligible for well measured decays where the charmed 
meson is almost fully reconstructed, they can be very large in other cases.
Parametrising carefully such corrections as a function of the event topology
is mandatory to achieve a good sensitivity at high frequency. That is valid,
to some extent, also for $\Ds \ell$ analyses, although the variety of
topologies is more limited.
These considerations suggest that a big factor could still be gained 
in the statistical power of the DELPHI analyses at high 
frequency\footnote{In the DELPHI $\Ds \ell$ analysis the fit error
is not used to estimate event-by-event the decay length uncertainty: the
average resolution is taken instead.}.

In Fig.~\ref{fig:comp2}b, the two best analyses from SLD, the inclusive lepton
and fully inclusive analyses, are compared with the corresponding best 
available at LEP. Compared to the ALEPH analysis, the SLD inclusive lepton
analysis at low frequency has errors larger by more than a factor of two,
due to the smaller statistics, only partially compensated by the better flavour tag at production time (see Section~\ref{sec:incll}). As the frequency
increases, the decay length resolution becomes more and more relevant, 
and the two analyses become eventually equivalent.

The comparison of the SLD and DELPHI fully inclusive analyses shows
the difficulty of such a method at LEP.

\begin{figure}[bt!]
 \vspace{6.2cm}
  \includegraphics{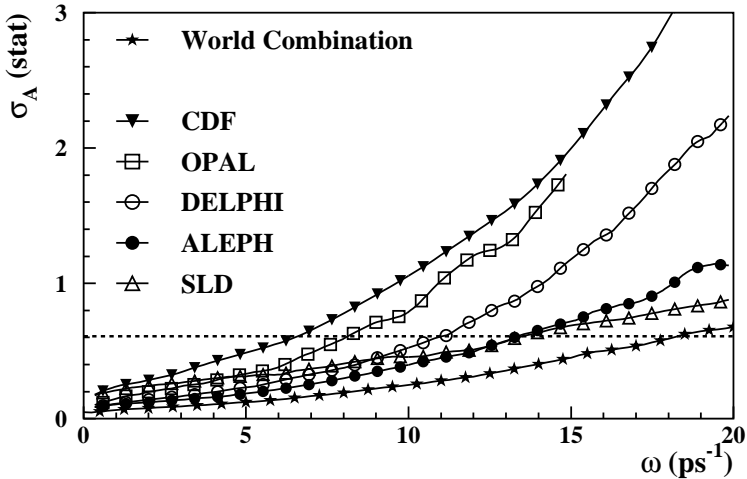}
 \caption{\it
   Comparison of the statistical errors on the amplitude for the combination
of analyses experiment by experiment. At high frequency the world average is 
dominated by SLD and ALEPH. 
    \label{fig:compexp} }
\end{figure}

In Fig.~\ref{fig:compexp} the combinations of the analyses 
from each experiment are compared. At high frequency the 
world combination is dominated by SLD and ALEPH. 

\subsection{Results from the world combination}

\begin{figure}[b!]
 \vspace{11.2cm}
  \includegraphics{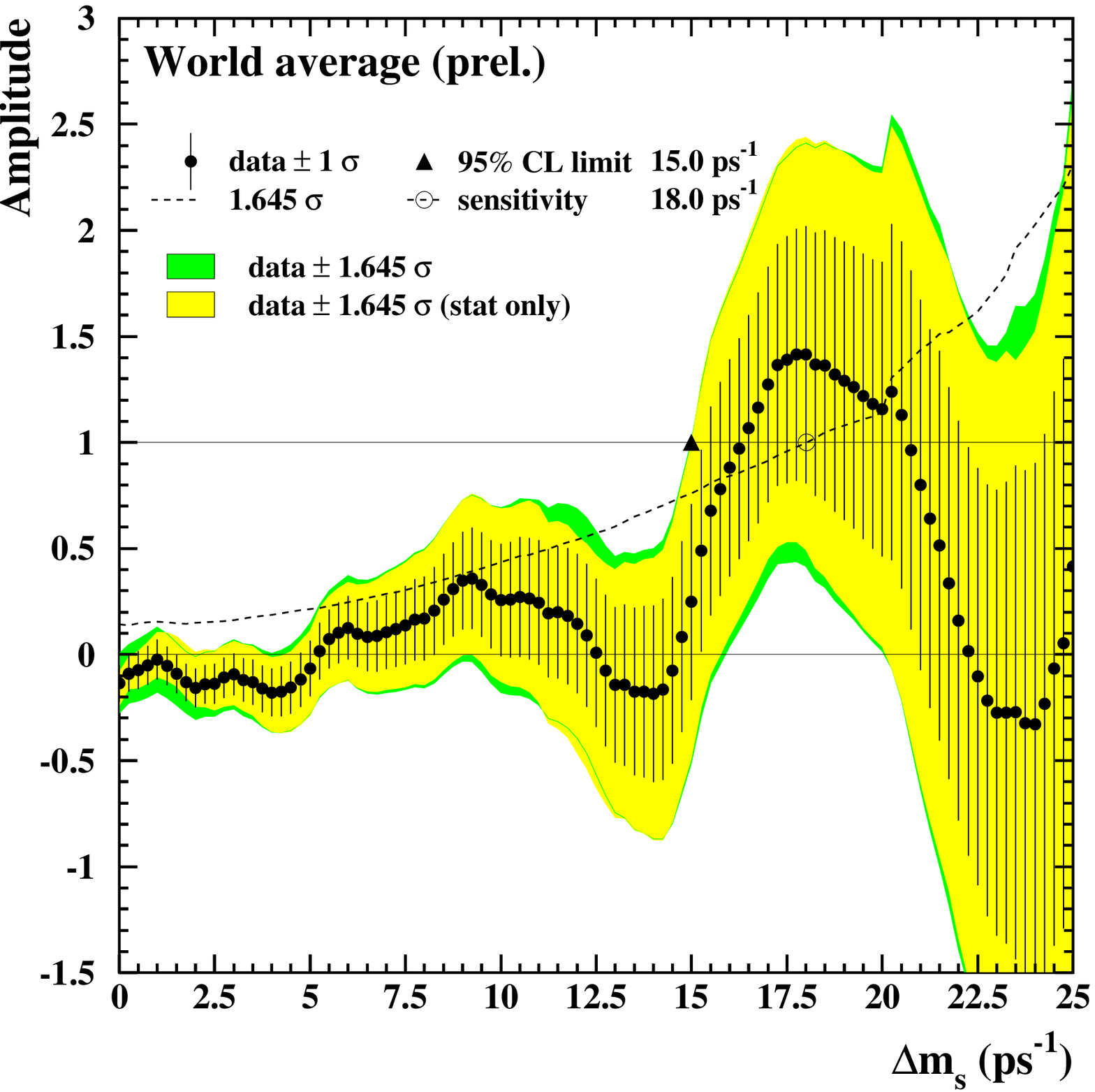}
 \caption{\it
      Measured amplitude as a function of test frequency 
for the world combination.
    \label{fig:ampw} }
\end{figure}
The amplitude spectrum for the world combination, 
with statistical and systematic errors is shown in Fig.~\ref{fig:ampw}.
A lower limit of \mbox{$\Dms>15.0 \ \ips$} is derived, while the
expected limit (sensitivity) is \mbox{$\Dms>18.0 \ \ips$}.
The difference is due to the positive amplitude values measured around 
\mbox{$17 \ \ips$}, compatible with one, as expected in the presence
of signal. The error on the amplitude at high frequency 
(\mbox{$\Dms\approx 20 \ \ips$}) is reduced by about a factor of two
compared to the world combination of Summer 1999, mostly because
of the progress of the SLD and ALEPH analyses.

\begin{figure}[tb!]
  \vspace{6.8cm}
  \includegraphics{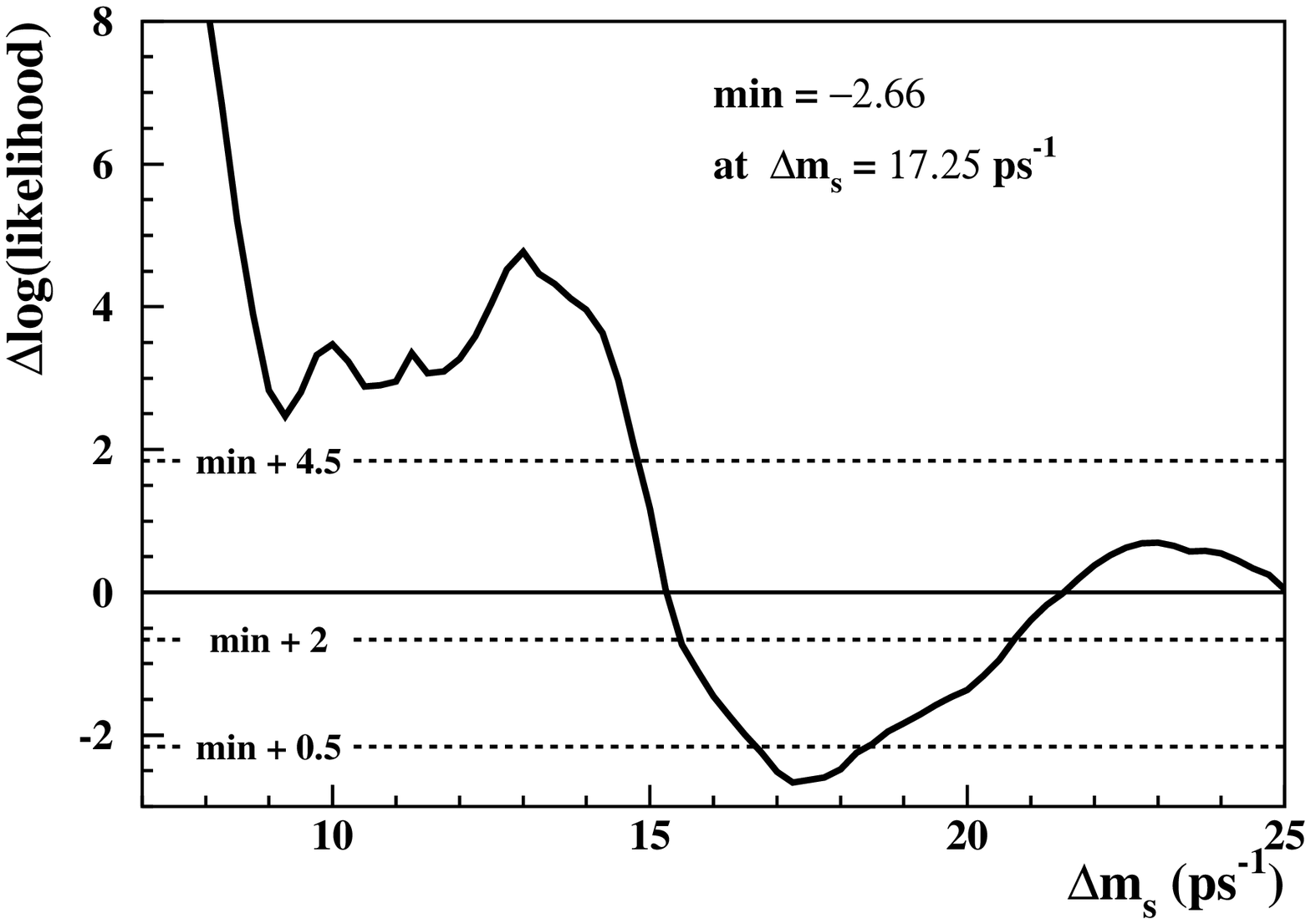}
  \caption{\it
    Likelihood profile as derived from 
the amplitude spectrum (world combination). The dashed lines would represent the $1-2-3\ \sigma$ levels, if the likelihood was parabolic in a range wide
enough around the minimum.
    \label{fig:like} }
\end{figure}

The amplitude spectrum can be translated to a likelihood profile,
referred to the asymptotic value for 
\mbox{$\Dms \to \infty$} (see Fig.\ref{fig:like}). A minimum
is observed at $\Dms \approx 17\ \ips$. The deviation of the measured
amplitude from $\amp = 0$ around the likelihood minimum 
is about $2.4 \sigma$. Such a value cannot be used
to assess the probability of a fluctuation,
since it is chosen {\em a posteriori} among all the points
of the frequency scan performed. On the other hand because the amplitude
measurements at different frequencies are correlated, the probability
of observing a minimum as or more incompatible
with the hypothesis of background than the one found in the data,
needs to be estimated with toy experiments\cite{noi}. 
Such a probability is found to be about $3\%$.

\begin{figure}[tb!]
 \vspace{8.5cm}
  \includegraphics{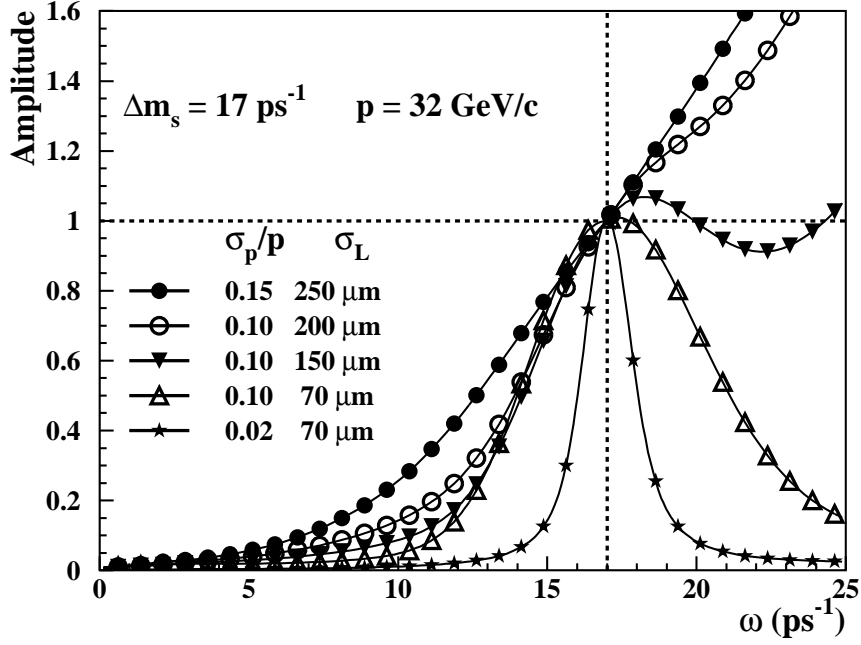}
 \caption{\it
   Expected amplitude shape for a true frequency $\Dms = 17 \ \ips$, monochromatic $\Bs$ mesons of $p=32\ \gevc$ and different values of momentum and decay length resolutions (taken to be Gaussian).
    \label{fig:ampl} }
\end{figure}

An interesting issue is to which extent the observation is compatible
with the hypothesis of signal. This cannot be assessed quantitatively in
 a non-trivial way. In Fig.~\ref{fig:ampl} the expected amplitude shapes,
calculated analytically\cite{noi},
are shown for the simple case of monochromatic  $\Bs$ mesons of  $p=32\ \gevc$
which oscillates with a frequency of  $17 \ \ips$, with different
(Gaussian) resolutions in momentum and decay length. The shapes 
are largely different; the only solid features are that the expectation is
$\amp=1$ at the true frequency and $\amp=0$ far below the true frequency. 

In the world combination many analyses contribute, which have very different
momentum and decay length resolution. In the most
sensitive analyses, even, each event enters with its specific
estimated resolutions, therefore contributing 
with a different expected amplitude shape.
Calculating the expected amplitude shape for the world combination 
in the hypothesis of signal is therefore, at the moment, completely impractical.

It can certainly be stated, however, that qualitatively the shape observed
in Fig.~\ref{fig:ampw} is well compatible with the hypothesis of a signal
at $\Dms\approx 17\ \ips$.

\subsection{Indirect constraints}

Indirect constraints on $\Dms$ can be derived, within the 
Standard Model framework, from other physics quantities.

\begin{figure}[tb!]
 \vspace{4.5cm}
  \includegraphics{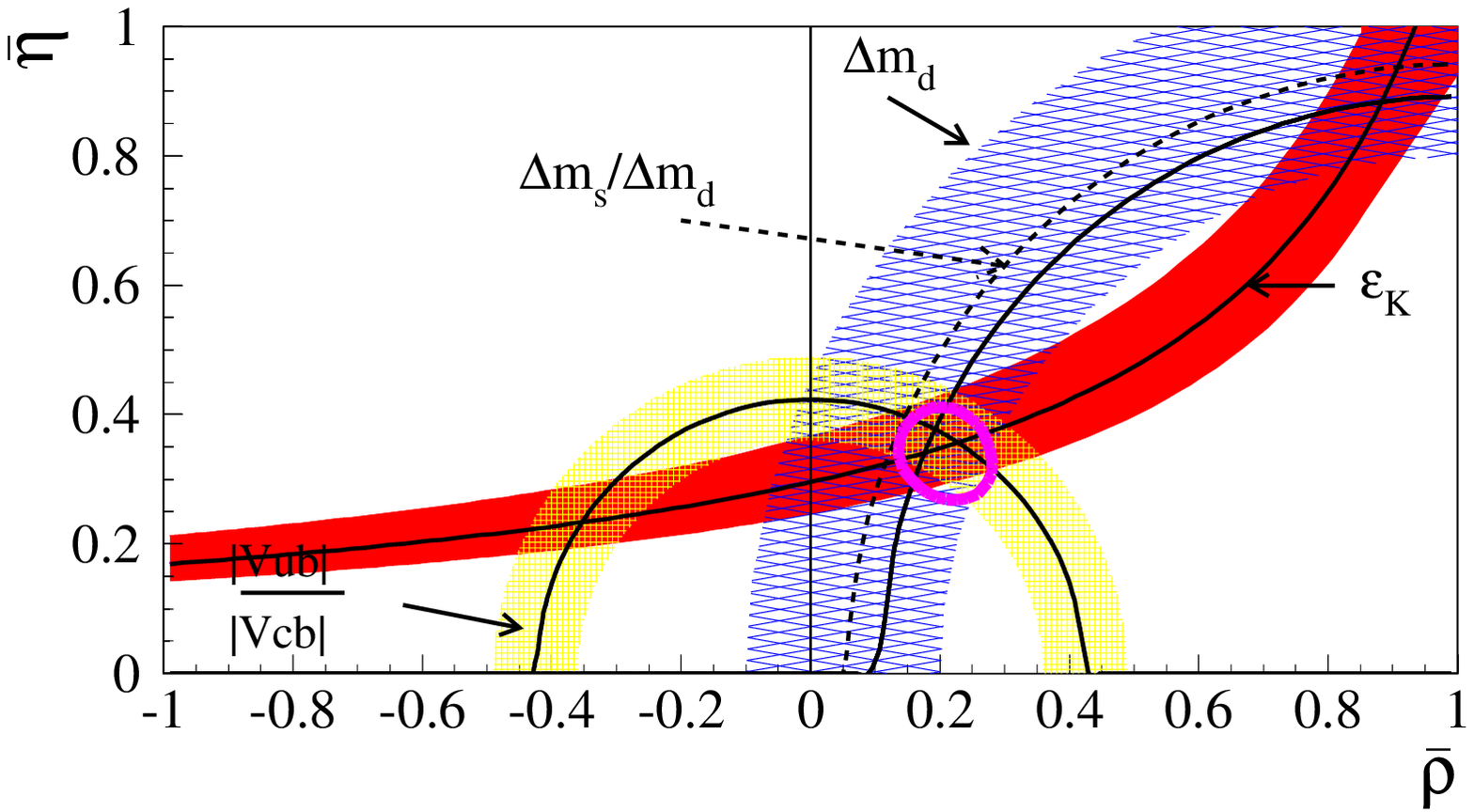}
  \includegraphics{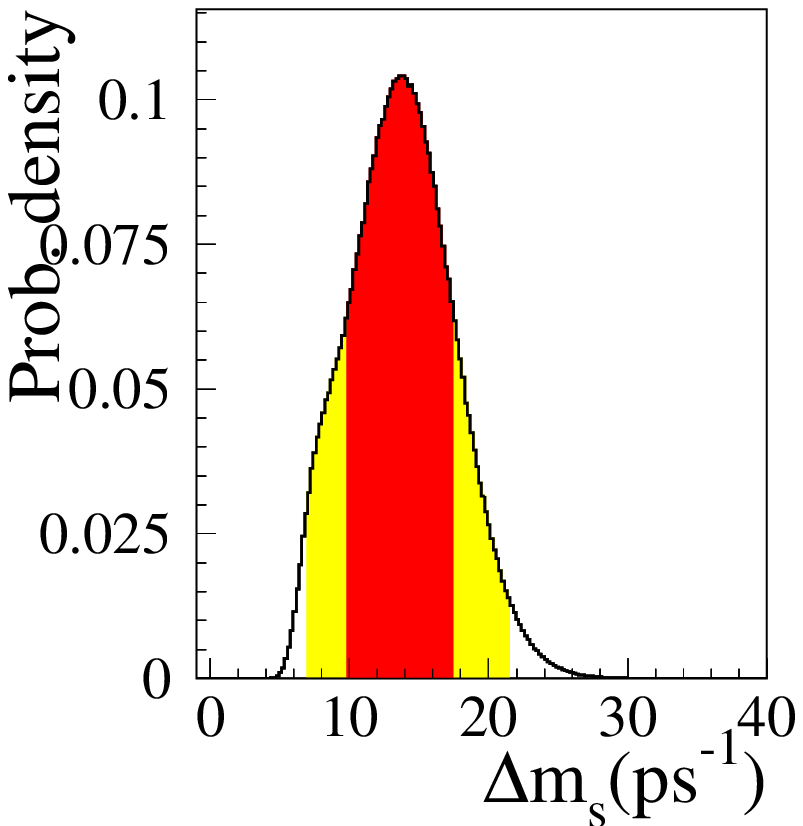}
 \caption{\it
   Constraints in the ($\rho,\eta$) plane from charmless $b$ decays, CP
violation in the kaon system, and $\B$ meson oscillations (a). Constraint
on the $\Dms$ derived from the fit with the
$\Bs$ oscillation information removed (b).
    \label{fig:stocchi} }
\end{figure}

Measurements of charmless $b$ decays, CP
violation in the kaon system, and $\B$ meson oscillations
can all be translated, with nontrivial theoretical input,
to constraints on the ($\rho,\eta$) parameters\cite{stocchi}, and combined, 
as shown in Fig.~\ref{fig:stocchi}a.
If the limit on $\Dms$
is removed from the fit, a probability
density function can be extracted from the other parameters,
shown in Fig.~\ref{fig:stocchi}b. The preferred value is
\mbox{$\Dms = 14.9 ^{+4.0} _{-3.6} \ \ips$}, perfectly compatible
with the indication observed in the combination of $\Dms$ analyses.

The present world average of the width difference in the $\Bs$ 
system\cite{roudeau}, 
\mbox{$\Delta \Gamma_s /  \Gamma_s = 0.16\, ^{+0.08} _{-0.09}$}, 
can be also translated to a value
for the oscillation frequency, using the prediction of NLO+lattice
calculations\cite{beneke} for the ratio $\Delta \Gamma_s / \Dms$. That gives 
\mbox{$\Dms = 16 \, ^{+8} _{-9} \pm 5\ \ips$}, very similar to the 
previous result, but with larger errors.

\section{Conclusions}

The $\Bd$ oscillation frequency has been precisely measured by CDF, SLD and the LEP eperiments:
\begin{equation}
\Dmd = 0.497 \pm 0.014 \ \ips \ .
\end{equation}
Asymmetric $\B$ factories will soon improve substantially on such result.

$\Bs$ oscillations are not yet resolved. A lower limit is set
\begin{equation}
\Dms > 15.0 \ \ips \ \ \ \ \ @ \ 95\% \ {\mathrm{C.L.}} \ \ .
\end{equation}
The expected limit is $18.0 \ \ips$, the difference being due to a deviation
of about $2.4\sigma$ from $\amp=0$ observed for frequencies around $17 \ \ips$.
The deviation is qualitatively compatible with the pattern expected for
a signal. The probability that it be due to a fluctuation is estimated to be about $3\%$.

\section{Acknowledgements}
I wish to thank the organisers of the Conference for inviting me, and
in particular Alberto Reis, amicable host.
I am grateful to Ga\"{e}lle Boix and Achille Stocchi, 
who provided several plots
and numbers.

\end{document}